\documentstyle[twocolumn,prl,aps]{revtex}
\begin{document}
\draft

\title{
Valence band structure, edge states and interband absorption
in Quantum Well Wires in high magnetic fields
}

\author{G. Goldoni}

\address{Departement Natuurkunde, Universiteit Antwerpen (UIA),
Universiteitsplein 1,
B-2610 Antwerpen, Belgium \\
and Istituto Nazionale Fisica della
Materia, Dipartimento di Fisica, Universit\`a di Modena, \\
Via Campi 213/A, I-41100 Modena, Italy}

\author{A. Fasolino}

\address{
Istituto Nazionale Fisica della
Materia, Dipartimento di Fisica, Universit\`a di Modena,  \\
Via Campi 213/A, I-41100 Modena, Italy}

\date{\today}
\maketitle

\begin{abstract}
We present a theoretical study of the magnetic band structure
of conduction and valence states in Quantum Well Wires in high
magnetic fields. We show that hole mixing
results in a very complex behavior of valence edge states with respect
to conduction states, a fact which is likely to be important in
magneto-transport in the Quantum Hall regime.
We show how the transition from one-dimensional subbands to edge states
and to Landau levels can be followed by optical experiments
by choosing the appropriate, linear or circular, polarization of the light,
yielding information on the one-dimensional confinement.
\end{abstract}

\pacs{}

Quasi-one-dimensional (Q1D) semiconductor structures are promising
systems for the investigation of novel optical and transport
properties, as well as for potential technological
applications.~\cite{cingo} The nature of {\em magnetic} states in
Q1D structures is of primary importance in the understanding of
transport properties, e.g., in the Quantum Hall regime.  Besides,
characterization of the currently challenging growth of these
structures and informations on the lateral confinement can be
gained by use of high magnetic fields; in fact, in Q1D structures,
the magnetic field provides an additional tunable confinement
length which can range from much larger to much smaller than the
lateral dimension of the wires and can therefore be used to
characterize the lateral confinement length in these structures.

Optical characterization of Q1D structures provides very detailed
information on electronic states; band structure effects, such as
the coupling between heavy-hole (HH) and light-hole (LH) valence
states, have been shown to affect the interband optical absorption
of Q1D and of laterally modulated structures at zero and low
magnetic field,~\cite{Bockelmann92} giving rise to anisotropic
absorption for linearly polarized light.

In this paper we present a theoretical study of the electronic and
optical properties up to high magnetic fields of Quantum Well
Wires (QWW) obtained by lateral etching of a Quantum Well
(QW).~\cite{marzin} The alternative method of growth of Q1D
structures by deposition on non-planar substrates is also used to
achieve smaller confinement lengths.~\cite{kapon,rinaldi} In QWWs,
electronic states have either a quasi-2D (Q2D) or a Q1D character,
depending on the size of the additional lateral confinement with
respect to the underlying QW.

As mentioned above, theoretical \cite{Bockelmann92,citrin} and
experimental \cite{rinaldi,kohl,tanaka,notomi} investigations have
shown that, at zero and low magnetic field, the in-plane
anisotropy induced by 1D quantization can be probed by linearly
polarized interband spectroscopy. Here we show that, in high
magnetic fields, instead, the anisotropic absorption for
circularly polarized light of different helicity is more
appropriate to study the features resulting from the field-induced
hole-mixing; A comparison of linearly and circularly polarized
light absorption as a function of the field directly probes the
lateral potential by monitoring the transition from the low field
regime, dominated by 1D confinement, to the high field regime,
dominated by Landau quantization.

In Sec.~\ref{sec:method} we outline the method of calculation of
conduction and valence states in Q1D structures from zero to high
magnetic fields. In Sec.~\ref{sec:edgestates} we focus on the
different nature of valence and conduction edge states in QWWs. In
Sec.~\ref{sec:optics} we present the calculated optical absorption
spectra in linearly and circularly polarized light. Finally,
summary and conclusions are given in Sec.~\ref{sec:conclusions}.

\section{method}
\label{sec:method}

We consider a QWW such as shown in Fig.~1, which we model by a
(001)-grown QW, of length $d_z$, and a lateral infinite square
well along (100), of length $d_x$. In the following we call
$V^e(z), V^h(z) $ the QW confining potential for electrons and
holes, respectively, and $V( x)$ the infinite well potential
acting laterally on both electrons and holes. We will also
consider this system in presence of a magnetic field B directed along the
(001) direction. Previous calculations of hole states in $B\ne0$
have considered weak sinusoidal potentials, which allow a simpler
analytical approach.~\cite{Bockelmann92}
  For both the B=0 and B$\neq$0 cases we will
also calculate the absorption strength for dipolar interband
transitions for linearly and circularly polarized light
perpendicular to the layers, as illustrated in Fig.~1. As we will
show below, the assumption of an infinite well in the
$x$-direction
 allows to reduce the 2D  problem to two
1D  problems.  In other words, it allows to factorize
the electron envelope function $\Psi^e({\bf r})$ into the product
of the envelope functions of the QW bound states $\psi^e(z)$ times
the envelope functions $\phi(x)$ of the lateral potential $V(x)$,
also in presence of a magnetic field. Furthermore, the same
factorization applied to the uncoupled HH and LH states serves as
a suitable representation to include the hole mixing. Conversely,
the choice of a QW profile in the $z$ direction can be easily
extended to more complex potential profiles, such as coupled QWs.

We obtain the electron and hole
energy levels and envelope functions by
solving separately the effective-mass electron and hole Hamiltonians.

\subsection{Case $B=0$}

In zero field the effective-mass electron Hamiltonian reads
\begin{equation}
H^e = -\frac{\hbar^2}{2}
\frac{\partial}{\partial z}\frac{1}{m^*(z)}\frac{\partial}{\partial z}
+\frac{\hbar^2}{2m^*(z)}
\left[ -\frac{\partial^2}{\partial x^2} + k_y^2 \right]
+ V(x) + V^e(z) \,,
\label{He}
\end{equation}
where $m^*(z)$ is the bulk conduction electron effective mass
which depend on the material and, therefore, on the $z$
coordinate; the usual current-conserving kinetic
operator~\cite{BenDaniel}
has been used accordingly. We neglect non-parabolicity effects.

The hole Hamiltonian is
\begin{equation}
{\bf H}^h= \bordermatrix{&+3/2&-1/2&+1/2&-3/2 \cr
     & H_h & c_+ & b_+ & 0 \cr
     & c_- & H_l & 0 & -b_+ \cr
     & b_- & 0 & H_l & c_+ \cr
     & 0 & -b_- & c_- & H_h \cr
     } + \left[V(x) + V^h(z)\right] {\bf I}_4 \,,
\label{Hh}
\end{equation}
with
\begin{mathletters}
\begin{eqnarray}
H_h & = & \frac{\hbar^2}{2m_0}
\left[\left(\gamma_1+\gamma_2\right)\left(\frac{\partial^2}
{\partial x^2}-k_y^2\right)
+ \frac{\partial}{\partial z}
\left(\gamma_1-2\gamma_2\right)
\frac{\partial}{\partial z} \right]
\,, \\
H_l & = & \frac{\hbar^2}{2m_0}
\left[\left(\gamma_1-\gamma_2\right)\left(\frac{\partial^2}{\partial
x^2}-k_y^2\right)
+ \frac{\partial}{\partial z}
\left(\gamma_1+2\gamma_2\right)
\frac{\partial}{\partial z} \right]
\,, \\
c_\pm & = & -\sqrt{3}\frac{\hbar^2}{2m_0}
\left[ \gamma_2 \left(\frac{\partial^2}{\partial x^2}+k_y^2\right)
\pm 2 \gamma_3 k_y \frac{\partial}{\partial x}\right]
\,,\\
b_\pm & = & -\sqrt{3}
\frac{\hbar^2}{2m_0}
\left( \frac{\partial}{\partial x}\pm k_y\right)
\left(\gamma_3\frac{\partial}{\partial z}+
\frac{\partial}{\partial z}\gamma_3\right)
\,,
\end{eqnarray}
\end{mathletters}
where $m_0$ is the free-electron mass, $\gamma_1,\gamma_2$ and $\gamma_3$ are
the Luttinger parameters and ${\bf I}_4$ is the identity $4\times 4$ matrix.
Equation (\ref{Hh}) is obtained from the Luttinger
Hamiltonian~\cite{Luttinger56}
written in the basis of the eigenstates of the $J=3/2$ total angular momentum,
\begin{equation}
|\frac{3}{2},+\frac{3}{2}\rangle\,,
|\frac{3}{2},-\frac{1}{2}\rangle\,,
|\frac{3}{2},+\frac{1}{2}\rangle\,,
|\frac{3}{2},-\frac{3}{2}\rangle\,,
\label{Jbasis}
\end{equation}
after the substitutions $k_x\rightarrow -i\partial/\partial x,
 k_z\rightarrow -i\partial/\partial z $, and symmetrization of the
non-commuting products~\cite{bastard-wm} between
$\partial/\partial z$ and the $z$-dependent $\gamma_i$'s.

The eigenstates of $H^e$ can be factorized as $\Psi^e_{n,m} ({\bf
r})=e^{ik_yy} \phi_m(x) \psi_n^e(z)$, where $n$ labels the bound
states of the QW potential $V^e(z)$, and $m$ the analytical
solutions of the infinite well $V(x)$.  Consequently, the
eigenvalues of (\ref{He}) depend quadratically on the in-wire
momentum $k_y$.
 The solution of the valence states is more complex because the
off-diagonal terms in ${\bf H}^h$ couple the solutions of the
diagonal, electron-like terms $H_h$ and $H_l$.  Following
Bockelmann and Bastard,~\cite{Bockelmann92} we first solve the
electron-like HH and LH Hamiltonians, $H_h+V^h(z)+V(x)$ and
$H_l+V^h(z)+V(x)$, giving the factorized solutions
$\Psi^{HH}_{n_h,m}=e^{ik_yy}\psi_{n_h}^{HH}(z)\phi_m(x) $ and
$\Psi^{LH}_{n_l,m}=e^{ik_yy}\psi_{n_l}^{LH}(z)\phi_m(x) $, where
$n_h, n_l$ label the HH and LH states. Then we expand the solution
of the hole Hamiltonian ${\bf H}^h$ onto the following basis set
\begin{mathletters} \begin{eqnarray} |HH^+;n_h,m\rangle & = &
(\Psi^{HH}_{n_h,m},0,0,0), \\ |LH^-;n_l,m\rangle & = &
(0,\Psi^{LH}_{n_l,m},0,0), \\ |LH^+;n_l,m\rangle & = &
(0,0,\Psi^{LH}_{n_l,m},0), \\ |HH^-;n_h,m\rangle & = &
(0,0,0,\Psi^{HH}_{n_h,m}). \end{eqnarray} \end{mathletters} In the
expansion, we let $n_h,n_l$ run over all $N_h$ bound HH and $N_l$
bound LH states of $V^h(z)$, while $m$ runs over $M$ eigenstates
of $V(x)$ (typically we take $M=15$). Finally we diagonalize the
resulting $2 \times (N_h+N_l) \times M $ Hamiltonian. Due to the
off-diagonal couplings, therefore, the valence band states cannot
be assigned to a single set of indices $(n,m)$, in contrast to
conduction electrons; this mixing results in a strongly
non-parabolic in-wire energy dispersion.

When we diagonalize the hole Hamiltonian ${\bf H}^h$,
we do not make the axial approximation,
as assumed in Ref.~\onlinecite{Bockelmann92}, which is more
severe in QWWs than in QWs as shown later in Section III.

\subsection{Case $B\neq0$}

When we include a uniform magnetic field
$\mbox{\boldmath $B$}=(0,0,B)$ along the
QW growth direction, described by the vector
potential $\mbox{\boldmath $A$}=(0,Bx,0)$,
the electron Hamiltonian $H^e$, neglecting
the electron spin splitting, reads
\begin{equation}
-\frac{\hbar^2}{2}
 \frac{\partial}{\partial z} \frac{1}{m^*(z)} \frac{\partial}{\partial z}+
\frac{1}{m^*(z)/m_0}\left[-\frac{\hbar^2}{2m_0} \frac{\partial^2}{\partial
x^2}+ \frac{m_0\omega^2}{2}\left(x-x_0\right)^2\right] + V^e(z) + V(x)
\end{equation}
and the hole Hamiltonian ${\bf H}^h$ becomes
\begin{equation}
{\bf H}^h -2\kappa\mu_BB{\bf J}_z\,,
\end{equation}
where ${\bf H}^h$ is given by (\ref{Hh}) with
\begin{mathletters}
\begin{eqnarray}
H_h & = &
\left(\gamma_1+\gamma_2\right)
\left[\frac{\hbar^2}{2m_0} \frac{\partial^2}{\partial x^2}-
\frac{m_0\omega^2}{2}\left(x-x_0\right)^2\right]  +
\frac{\hbar^2}{2m_0} \frac{\partial}{\partial z}
\left(\gamma_1-2\gamma_2\right)
\frac{\partial}{\partial z}
+ V^h(z)+V(x)\,, \\
H_l & = &
\left(\gamma_1-\gamma_2\right)
\left[\frac{\hbar^2}{2m_0} \frac{\partial^2}{\partial x^2}-
\frac{m_0\omega^2}{2}\left(x-x_0\right)^2\right]  +
\frac{\hbar^2}{2m_0} \frac{\partial}{\partial z}
\left(\gamma_1+2\gamma_2\right)
\frac{\partial}{\partial z}
+ V^h(z)+V(x)\,, \\
c_\pm & = &
-\sqrt{3}\left[\gamma_2\left(\frac{\hbar^2}{2m_0}\frac{\partial^2}{\partial
x^2}+\frac{m_0\omega^2}
{2}\left(x-x_0\right)^2\right)\pm\gamma_3\frac{\hbar\omega}{2}
\left(1+2\left(x-x_0\right)\frac{\partial}{\partial x}\right)\right] \,,\\
b_\pm & = & -\sqrt{3}
\left[\frac{\hbar^2}{2m_0}
\frac{\partial}{\partial x}\pm\frac{\hbar\omega}{2}\left(x-x_0\right)\right]
\left(\gamma_3\frac{\partial}{\partial z}+
\frac{\partial}{\partial z}\gamma_3\right)
\,.
\end{eqnarray}
\end{mathletters}
${\bf J}_z$ is the diagonal matrix representing the $z$-component of
the angular momentum operator in the basis (\ref{Jbasis});
$\kappa $ is an additional Luttinger parameter,
$\omega=e B/m_0$ is  the cyclotron frequency and
$x_0=-l_m^2k_y$ is the semiclassical `orbit center', where
$l_m=(\hbar/eB)^{1/2}$ is the magnetic length. The hole eigenstate
dependence on the in-wire wavevector $k_y$ of the $B=0$ case, is now
replaced by the dependence on $x_0$.

Both for $H^e$ and for the diagonal terms of ${\bf H}^h$ the
lateral potential in the $x$ direction is, at $B\neq0$, the sum of
the infinite well plus a parabolic effective potential. The latter
term is equal for the two Hamiltonians so that, at each $B$ and
each $x_0$, the electron and hole $\phi_m(x)$'s are the same, as
in the $B=0$ case.  In high magnetic fields, the $\phi_m(x)$'s
reduce to harmonic oscillator eigenstates when $x_0$ is far from
the barrier, i.e., when the confinement along $x$ due the
parabolic magnetic potential is not affected by the infinite
barriers; instead, when $x_0$ is near the barriers the
$\phi_m(x)$'s are the so-called edge states.~\cite{MacDonald84} In
view of the possible application to other confinement potentials,
we have found convenient to compute both the $\psi_n(z)$'s and the
$\phi_m(x)$'s numerically by direct integration of the
Schr\"odinger equation in real space.

\section{Energy levels and edge states of QWWs}
\label{sec:edgestates}

In Fig.~2(a) and Fig.~2(b) we show the calculated hole subbands for
QWWs formed by a GaAs/Al$_{0.35}$Ga$_{0.65}$As QW of width
$d_z=10\,\mbox{nm}$ and by lateral infinite wells of width
$d_x=30\,\mbox{nm}$ and $d_x=100\,\mbox{nm}$ respectively. Each HH
and LH bound state of the QW is split by the lateral potential
into a series of 1D subbands of mixed character. In Table~I we
give the orbital composition of selected states as compared to the
purely HH or LH states of the QW at zero in-plane wavevector. As
expected, the additional confinement energy and hole mixing is
much less for $d_x=100\,\mbox{nm}$ than for $d_x=30\,\mbox{nm}$.
Furthermore, in Fig.~2(a) and Table~I
we compare the calculated 1D hole subbands
with those obtained within the axial approximation for the same
structure.  The anisotropy-induced couplings have a rather strong
effect on the curvature of the subbands; in particular, the
electron-like curvature of the
light hole derived 1D subbands is strongly reduced in the axial
approximation.  In fact, as sketched in the inset of Fig~2(b),
due to the quantization of the lateral
wavevector $k_x$, each in-wire momentum $k_y$ corresponds to an
effective in-plane wavevector ${\bf
k}_\parallel=(k^{\mbox{\small eff}}_x,k_y)$, away from the (010)
direction, of the underlying, anisotropic QW band
structure.~\cite{Altarelli85}  Here
$k^{\mbox{\small eff}}_x=m\pi/d_x $, where $m$ is the QWW subband index.
Therefore, as it is apparent in Fig.~2(a), in QWWs not only the in-wire
dispersion but also the hole confinement energies are affected by
the axial approximation.

In Fig. 3 we show the electron and hole magnetic levels at
$B=10\,\mbox{T}$
for the same QWW of Fig.~2(b). We plot the magnetic levels as a
function of the semiclassical `orbit center' $x_0$.  The most
striking effect is that, while the electrons (Fig.~3(a)) display a
ladder of edge states with well defined increasing oscillator
number,~\cite{buttiker} the valence edge states (Fig.~3(b)) have
complicated shapes which are likely to be of relevance for
magneto-transport, particularly in the integer Quantum Hall regime.  As
for the 1D subbands, also here the axial approximation is very
severe.

The comparison of Figs.~3(a) and 3(b) clearly shows that, for holes, the
cross over from bulk Landau levels to edge states close to the
barrier is less straightforward than for electrons. In fact, each
hole state is a mixture of oscillator states of different quantum
number $m$.  As a consequence, each component starts feeling the
effect of the barrier at a different value of the magnetic field.
As shown in Table II, the second hole level is the only one almost
purely $m=0$ and, indeed, it starts deviating from a flat dispersion
at the same $x_0$ as the $m=0$ electron state, while all others
have a mixed composition resulting in more complicated
dispersions.

This point is further clarified by aid of Fig.~4 and Table II.
In Fig.~4 we compare the calculated hole magnetic levels for the
two QWWs of Fig.~2 ($d_z=10\,\mbox{nm}$ and either
$d_x=30\,\mbox{nm}$ or $d_x=100\,\mbox{nm}$) in a magnetic field of
$10\,\mbox{T}$ and $30\,\mbox{T}$.  We recall that at
$B=10\,\mbox{T}$, ~$l_m \simeq
8\,\mbox{nm}$, i.e., comparable to half the thinner
$d_x$, while at $B=30\,\mbox{T}$, $l_m \simeq 4.7\,\mbox{nm}$, i.e.,
smaller than half  $d_x$ for both structures.  Therefore,
at $B=10\,\mbox{T}$, for $d_x=30\,\mbox{nm}$ (panel~4(a)) all Landau
levels are
perturbed by the lateral barriers also at the center of the well,
while for $d_x=100\,\mbox{nm}$ (panel 4(c)) the first edge states
evolve to flat Landau levels away from the barrier. Conversely, at
$B=30\,\mbox{T}$ (panels 4(b), 4(d)) we find that the energy structure
of all edge states down to -40 meV is the same for the two values of
$d_x$ and is dominated by Landau
quantization, due to the field, rather than
by the one-dimensional confinement.

Table II gives a rationale for
this behavior. It can be seen that very few states are
basically composed of only one harmonic oscillator number and that the
mixing increases with increasing B.
Furthermore, at high fields, the eigenvalues
at the center of the well of the low oscillator index levels do
not depend any more on the lateral potential. This transition
takes place at different values of the magnetic field for each
level as illustrated in Fig.~5. Here we show the evolution of
electron and hole states as a function of the field for two values
of $d_x$. The striking difference between electron and hole levels is
related to the mixed composition of the hole states, given in Table II;
The non-trivial behavior of hole levels in Fig.~5 suggests that
the transition from the low field regime,
dominated by 1D quantization, to the high field regime, dominated
by Landau quantization, is associated with large changes
in the wavefunction composition, with consequent influence on the
matrix elements for optical transitions, which we calculate in the next
section.
Our aim is to show that the transition from the low to the high
field regime can be monitored by optical experiments with linearly
and circularly polarized light, giving unambiguous information on
the lateral confinement.

\section{Optical absorption in QWW}
\label{sec:optics}

At B=0, the two in-plane linear polarizations of the light induce
different inter-band absorption probability, as a consequence of the
one-dimensionality of the hole states,~\cite{Bockelmann92} while the two
circular polarizations give the same absorption intensity. At the opposite
limit of very large fields, the situation is reversed: we expect the two
in-plane linear polarizations to give degenerate spectra, due to the
recovered two-dimensional character of hole states, while the two circular
polarizations of the light, being coupled to different components of the
spin-orbit coupled states, give different spectra. Therefore, in the
intermediate
regime, we expect large changes in the absorption spectra for both
circularly and linearly polarized light, as shown next.

Strictly speaking, a calculation of magneto-optical properties in Q1D
structures should include excitonic effects. The exact inclusion of the
Coulomb interaction is, however, a complicated task in itself and
theoretical calculations exist only for model, purely 1D structures in
zero field.~\cite{ogawa} However, several studies~\cite{ancil,goldoni}
have shown that, in Q2D structures, a perturbative excitonic
correction~\cite{hasegawa}
added to the one-particle absorption spectra yield a good description of
magneto-optical experiments in the high field regime, provided the hole
mixing is included.  Therefore, we expect our results,
obtained in a one-particle approximation, to be
qualitatively correct, particularly in the high field regime close to the
transition to a 2D behaviour.

In Fig.~6, we compare the calculated optical absorption spectra for
linear and circular light polarization at $B=10\,\mbox{T}$ and
$B=30\,\mbox{T}$ and for $d_x=30\,\mbox{nm}$; we have labelled
the main features according to the orbital character of
the initial (hole) and final (electron) states involved in the
transitions.  The mixing of hole states makes that very many hole
levels have not vanishing oscillator strength for the transition
to the same electron state, making the spectra rich and
informative. For instance, at $10\,\mbox{T}$ the transition
$\mbox{LH}^-_{10}\rightarrow e^\downarrow_{10}$, which is induced
by $\sigma^+$ polarized light, gives rise to a single peak in the
spectra, while the transition $\mbox{LH}^+_{10}\rightarrow
e^\uparrow_{10}$ is split into three peaks in $\sigma^-$
polarization. In fact, as shown in Table II, there is only one
level (level 6) with strong LH$^-$ character and a strong $m=0$
component, while there are three levels with LH$^+$ and $m=0$
character (levels 5, 10 and a deeper one which is not reported in
the table).

Figure 6 shows that, while the spectra for
light linearly polarized along the two in-plane direction of the
QWW become very similar at high fields, as we expect from the
isotropic character of the orbital motion, large differences are
present in the spectra calculated for the two circular
polarization of the light, as observed in Q2D
structures.~\cite{ancil}  Hence, these differences can be regarded
as a fingerprint of the transition from 1D  dynamics
to Landau quantization;  by studying the evolution of the linear
and circular absorption, the transition from the low field
to the high field regime can be followed, yielding quantitative
information on the effective length of the 1D confinement.

Figure 7 shows how the anisotropy between spectra obtained from
linearly polarized light decreases as a function of the field.
Calculations are performed for the same sample of Fig.~6.  The
anisotropy is expressed as $100\times (I_x-I_y)/(I_x+I_y)$, where
$I_x$ and $I_y$ are the heights of the lowest energy peak of the
absorption spectra calculated for the $x$ and $y$ polarizations,
respectively.  As expected from the above discussion, the
anisotropy is strongly quenched by the field and decreases
monotonously from $10\,\%$ at $B=0$ to $3\,\%$ at $B=20\,\mbox{T}$
for this sample. At fields larger than $20\,\mbox{T}$, this peak splits
into a double peak, as shown in Fig.~6 at $30\,\mbox{T}$, and the above
definition does not apply.

Finally, in Fig.~8 we show the evolution of the calculated spectra
for the two circular polarizations of the light as a function of
magnetic field for the $d_z=10\,\mbox{nm}$, $d_x=30\,\mbox{nm}$
sample. Hole mixing is responsible of complex features in the
low-energy range, which could hopefully be experimentally
detected.  In Fig.~8 the calculated spectra for the $\sigma^-$
polarization show a clear anticrossing behavior around
$1590\,\mbox{meV}$. This feature, which is due to HH-LH mixing and
would be absent in a simple, uncoupled-hole model, can be seen,
again, as a fingerprint of the transition between 1D subbands to
Landau levels of hole states. Note that this anticrossing is
totally absent in the corresponding peaks for the $\sigma^+$
polarization as a consequence of the smaller mixing of HH$^-$ and
LH$^-$ states as compared to HH$^+$ and LH$^+$ states (see Table II).

\section{conclusions}
\label{sec:conclusions}

We have presented a study of the electronic structure of rectangular QWWs
following the transition from the 1D quantization at zero magnetic fields
of the subband structure to Landau quantization at high magnetic fields.
We have focused on the behavior of the spin-orbit coupled valence band,
and, to this purpose, we have developed a method of solution of the
Luttinger Hamiltonian for Q1D structures both at B=0 and B$\neq$0. We have
shown that the hole mixing gives rise to a rather complex edge state
structure compared to electron states. Furthermore, we have shown that a
study of the optical interband transition for linearly and circularly
polarized light could be used to get informations on the lateral
confinement. As we discussed in Sec.~III, excitonic effects, which are
neglected in the present calculations, should not change qualitatively our
results, particularly in the regime of high fields. Furthermore, the
present one-particle calculation, which takes into account the coupled
nature of the valence subbands, is a necessary ingredient for a successive
calculation of the magneto-exciton.

 This study is also preliminary to a study of more
complex Q1D structures~\cite{kapon,rinaldi} showing very promising optical
properties.  In these structures, the two confinement directions are of
comparable width and have potential barriers of the same high and,
therefore, should be treated on the same foot. Theoretical investigations
of valence states in these structures are currently in progress. We hope
that the present work will stimulate further experimental and theoretical
work both on the transport and optical properties of these interesting
systems.

\acknowledgements

We thank F. Rossi for fruitful discussions.
We acknowledge the project ESPRIT 7260 (SOLDES)
for partial financial support. Part of this work has been carried out
within the European Human Capital \& Mobility Network magNET, n. ERBCHRXCT
920062 for

%
%
\begin{figure}
\caption{Sketch of the rectangular QWWs considered here.
We assume infinite potential barriers for the lateral well in the
$x$ direction. The direction of the magnetic field, as well as the
configuration for optical interband absorption with circularly and
linearly polarized light, are also indicated.}
\end{figure}

\begin{figure}
\caption{(a) One-dimensional valence subbands for a
GaAs/Al$_{0.35}$Ga$_{0.65}$As
QWW with $d_z$=10~nm and $d_x$=30~nm. Solid line: full inclusion of hole
mixing; dashed line: axial approximation. (b) One-dimensional valence subbands
for a GaAs/Al$_{0.35}$Ga$_{0.65}$As QWW with $d_z$=10~nm and $d_x$=100~nm.
The inset shows how the warped QW band structure is further quantized by
the one-dimensional potential so that
each in-wire momentum $k_y$ corresponds to an effective in-plane
wavevector ${\bf k}_\parallel=(k^{\mbox{eff}}_x,k_y)$,
away from the (010) direction, of the underlying, anisotropic QW band
structure.
}
\end{figure}

 \begin{figure}
 \caption{Conduction (a) and valence (b) edge state energy of a
GaAs/Al$_{0.35}$Ga$_{0.65}$As QWW with $d_z=10\,\mbox{nm}$ and
$d_x=100\,\mbox{nm}$ at $B=10\,\mbox{T}$ as a function of the
semiclassical `orbit center' $x_0$. The origin of energy is at the
conduction and valence bulk band edge, respectively.  The origin $x_0=0$
is chosen to coincide with the infinite barrier edge, indicated by the
vertical dotted line, the wire extending from $0$ to $-d_x$.  Conduction
(valence) edge states are bent upward (downward) due to magnetic
confinement close to the barrier.  Valence edge states calculated with
full inclusion of the hole mixing (solid line) are compared with those
calculated within the axial approximation (dashed line).}
 \end{figure}

\begin{figure}
 \caption{Valence edge states of
GaAs/Al$_{0.35}$Ga$_{0.65}$As QWWs with $d_z=10\,\mbox{nm}$: (a)
$d_x=30\,\mbox{nm}$, $B=10\,\mbox{T}$; (b) $d_x=30\,\mbox{nm}$,
$B=30\,\mbox{T}$; (c) $d_x=100\,\mbox{nm}$,
$B=10\,\mbox{T}$; (d) $d_x=100\,\mbox{nm}$, $B=30\,\mbox{T}$. $x_0=0$ is
set at the position of the right hand infinite barrier, indicated by
vertical dotted lines. Vertical dashed lines
in panels (a) and (b) indicate the wire center; the edge states are
symmetric with respect to the wire center. Notice
that, at $B=30\,\mbox{T}$, the edge states of the two samples have become
identical down to $\sim-40\,\mbox{meV}$.}
\end{figure}

\begin{figure}
 \caption{ Top panel: shift of the conduction edge states
calculated at the wire center as a function of the magnetic field for
$d_x=30$ nm (solid line) and $d_x=100$ nm (dashed line). Conduction edge
states subsequently converge to the bulk value.  Bottom panel: shift of
the valence edge states calculated at the wire center as a function of the
magnetic field for $d_x=30$ nm (solid line) and $d_x=100$ nm (dashed
line). The origin of energy is at the conduction and valence bulk band
edge, respectively. Notice that, contrary to electrons, the convergence to
bulk energies does not occur in a regular fashion for holes. }
\end{figure}

\begin{figure} \caption{ Calculated optical absorption spectra of
a GaAs/Al$_{0.35}$Ga$_{0.65}$As QWW with $d_z$=10~nm and
$d_x$=30~nm for linear (X, Y, Z) and circular ($\sigma^+$,
$\sigma^-$) light polarization at $B=10\,\mbox{T}$ and
$B=30\,\mbox{T}$, as indicated in each panel. Calculated spectra
have been
convoluted with a $5\,\mbox{meV}$ gaussian broadening. We have
labelled the main
transitions {\it initial state $\rightarrow$ final state}, where
the initial valence state is labelled according to its main
component (HH$^\pm_{nm}$ or LH$^\pm_{nm}$) and the final
electronic state is labelled as e$^{\uparrow \downarrow}_{nm}$ }
\end{figure}

\begin{figure}
\caption{Anisotropy of the calculated absorption spectra with
linearly polarized light as a function of the magnetic field and
for the sample of Fig.~6. $I_x$ and $I_y$ are defined in
Sec.~III.}
\end{figure}

\begin{figure}
\caption{Calculated optical absorption spectra of
a GaAs/Al$_{0.35}$Ga$_{0.65}$As QWW with $d_z=10\,\mbox{nm}$ and
$d_x=30\,\mbox{nm}$ for the two circular polarizations of the light.
Notice the avoided crossing around $E\sim1590\,\mbox{meV}$ for
$\sigma^-$, absent in the $\sigma^+$ polarization.}
\end{figure}


\begin{table}
\begin{tabular}{|c|c|ccc|cccccc|}
      &   QW  &
      \multicolumn{3}{c|}{ QWW, $d_x=100\,\mbox{nm}$ } &
      \multicolumn{6}{c|}{ QWW, $d_x=30\,\mbox{nm}$ }  \\
      & $d_x=\infty$  & \multicolumn{3}{c|}{ }   &
      \multicolumn{3}{c}{ Full calculation } &
      \multicolumn{3}{c|}{ Axial approximation }  \\\hline
      &  E         &  E &  HH/LH & $m=0 $
                   &  E &  HH/LH & $m=0 $
                   &  E &  HH/LH & $m=0 $         \\\hline
HH$_1$& -7.39  & -7.62  &  100  &  100 &   -9.94  &  98  &  98
& -10.14 & 98 & 99 \\
LH$_1$& -22.55 & -22.52 &   85  &   59 &   -23.06 &  84  &  81
& -23.41 & 89 & 88 \\
HH$_2$& -29.39 & -30.19 &   85  &   85 &   -35.89 &  67  &  62
& -35.64 & 63 & 57
\end{tabular}
\caption{Comparison of orbital character of the
QW bound states  and of the QWW levels at $k_y=0$ having mainly $m=0$
character at $B=0$ for the two differently confined  QWWs of Fig.~2.
For each QWW, we give the energy (in meV) of the lowest
level deriving from each pure QW state, the percentage of HH or LH
character, and the percentage of $m=0$ character.
The pure QW HH$_1$, LH$_1$ and HH$_2$ levels are
further confined by the QWW potential $V(z)$ and acquire  a mixed HH-LH
character.}
\end{table}
\begin{table}
\begin{tabular}{clllllllll}
\multicolumn{10}{c}{$d_z=10\,\mbox{nm}$, $d_x=30\,\mbox{nm}$,
$B=10\,\mbox{T}$ }\\\hline
  band & E(meV) & $HH_1^+$ & m & $HH_1^-$ & m   &  $LH_1^-$ & m
& $LH_1^+$ & m \\\hline
    1 & -10.24 & 0.89     & 0 &   -      & -   &   0.07    & 2
&   -    &  -   \\
    2 & -10.69 &  -       & - & 0.98     & 0   &    -      & -
&  0.01  & 2    \\
    3 & -16.29 & 0.72     & 1 &  -       & -   &   0.20    & 3
&   -    &  -   \\
    4 & -17.50 &  -       & - & 0.70     & 1   &    -      & -
&  0.20  & 1,3  \\
    5 & -19.35 &  -       & - & 0.36     & 2,0 &    -      & -
&  0.54  & 0,2  \\
    6 & -19.47 & 0.03     & 2 &  -       & -   &   0.82    & 0
&   -    &  -   \\
    7 & -21.09 &  -       & - & 0.28     & 3   &    -      & -
&  0.52  & 1    \\
    8 & -23.12 & 0.53     & 2 &  -       & -   &   0.36    & 2,4
&   -    &  -   \\
    9 & -24.01 & 0.09     & 3 &  -       & -   &   0.76    & 1
&   -    &  -   \\
   10 & -25.91 &  -       & - & 0.09     & 4   &   -       & -
&  0.78  & 0,2,4 \\\hline\hline
\multicolumn{10}{c}{ $d_z=10\,\mbox{nm}$, $d_x=30\,\mbox{nm}$,
$B=30\,\mbox{T}$ }
\\\hline
  band & E(meV) & $HH_1^+$ & m & $HH_1^-$ & m   &  $LH_1^-$ & m
& $LH_1^+$ & m \\\hline
     1 & -11.52 & 0.77     & 0 &   -      & -   &  0.17     & 2
&  -       & - \\
     2 & -15.96 &  -       & - &  0.95    & 0   &   -       & -
& 0.05     & 2 \\
     3 & -19.89 & 0.02     & 2 &   -      & -   &  0.85     & 0
&  -       & -  \\
     4 & -20.27 & 0.61     & 1 &   -      & -   &  0.29     & 3
&  -       & -  \\
     5 & -20.27 &  -       & - &  0.15    & 3   &   -       & -
& 0.64     & 1 \\
     6 & -20.78 &  -       & - &  0.24    & 2   &   -       & -
& 0.70     & 0 \\
     7 & -27.38 &  -       & - &  0.23    & 4   &   -       & -
& 0.64     & 2 \\
     8 & -28.42 & 0.51     & 2 &   -      & -   &  0.36     & 4
&  -       & - \\
     9 & -29.47 &  -       & - &  0.58    & 1   &   -       & -
& 0.25     & 3 \\
    10 & -30.50 & 0.07     & 3 &   -      & -   &  0.78     & 1
&  -       & -  \\ \hline\hline
\multicolumn{10}{c}{  $d_z=10\,\mbox{nm}$,
$d_x=100\,\mbox{nm}$,
$B=30\,\mbox{T}$ }\\\hline
  band & E(meV) & $HH_1^+$ & m & $HH_1^-$ & m   &  $LH_1^-$ & m
& $LH_1^+$ & m \\\hline
     1 & -11.56 & 0.76     & 0 &  -       & -   &  0.17     & 2
&   -      & -   \\
     2 & -15.93 &  -       & - & 0.95     & 0   &   -       & -
&  0.05    & 2 \\
     3 & -19.87 & 0.03     & 2 &  -       & -   &  0.84     &  0
&  -       & -  \\
     4 & -20.37 &  -       & - & 0.15     & 3   &   -       &  -
& 0.64     & 1 \\
     5 & -20.41 & 0.60     & 1 &  -       & -   &  0.29     & 3
&   -      & - \\
     6 & -20.79 &  -       & - & 0.24     & 2   &  -        & -
&  0.70    & 0 \\
     7 & -27.42 &  -       & - & 0.26     & 4   &   -       & -
& 0.61     & 2 \\
     8 & -28.54 & 0.49     & 2 &  -       & -   &  0.38     & 4
&  -       & -  \\
     9 & -29.37 &  -       & - & 0.54     &  1  &  -        & -
& 0.29     & 3  \\
    10 & -30.48 & 0.09     & 3 &  -       &  -  &  0.76     & 1
&  -       & -  \\ \end{tabular}
\caption{
 For different values of $d_z$, $d_x$ and $B$, this table gives, for the
lowest subbands, the energy and the $HH_1^+$, $HH_1^-$, $LH_1^-$, and
$LH_1^+$ character at $x_0$ located at the center of the lateral well.
For each component, the main Landau oscillator index is indicated. Notice
that at $B = 30$ T the eigenvalues with low oscillator index do not depend
on the confining length $d_x$.}
 \end{table}

\end{document}